\documentclass{ws-procs975x65}
\def\be{\begin{equation}}
\def\ee{\end{equation}}
\def\bea{\begin{eqnarray}}
\def\eea{\end{eqnarray}}
\def\l{\left}
\def\r{\right}

\begin{document}
\title{\uppercase{Apparent horizons in Clifton-Mota-Barrow inhomogeneous universe}}

\author{\uppercase{Vincenzo Vitagliano}$^1$, \uppercase{Valerio Faraoni}$^2$, \uppercase{Thomas P. Sotiriou}\\ and \uppercase{Stefano Liberati}$^3$}

\address{$^1$~CENTRA, Departamento de F\'isica, Instituto Superior T\'ecnico,
Universidade T\'ecnica de Lisboa - UTL, Av. Rovisco Pais 1, 1049 Lisboa, Portugal\\
$^2$~Physics Department and {\em STAR} Research Cluster, 
Bishop's University,\\
Sherbrooke, Qu\'ebec, Canada J1M~1Z7\\
$^3$~SISSA, Via Bonomea
265, 34136 Trieste, Italy and  \\INFN, Sezione di
Trieste, Italy
}

\begin{abstract}

We analyze the apparent horizon dynamics in the 
inhomogeneous Clifton-Mota-Barrow solution of Brans-Dicke 
theory. This solution models  a central matter 
configuration embedded in a cosmological background. In 
certain regions of the parameter space we find solutions 
exhibiting dynamical creation or merging of two horizons.

\end{abstract}

\bodymatter

\section{Introduction}

Scalar-tensor theories lead to a spacetime-dependent 
effective gravitational coupling  $G_{\textrm{eff}}$. In 
theories with this characteristic, the problem of understanding the behaviour of a local system embedded in a 
cosmological environment (cf. the McVittie 
solution\cite{McVittie:1933zz} of general relativity which 
is not yet well understood), is of particular interest. 
This motivated the study 
of inhomogeneous analytical solutions of scalar-tensor theories of gravity. 

The Clifton--Mota--Barrow spherically symmetric and time-dependent metric\cite{Clifton:2004st} is
\be\label{CMBmetric}
ds^2=
-A(\varrho)^{2\alpha}dt^2
+a^2(t) \l[\l(1+\frac{m}{2\alpha \varrho}\r)^{4} 
A(\varrho)^{\frac{2}{\alpha}( \alpha-1)(\alpha +2)}\r](d\varrho^2+\varrho^2d\Omega^2) \,,\nonumber
\ee
with 
\be
A(\varrho)=\frac{1-\frac{m}{2\alpha \varrho}}{1+\frac{m}{2 \alpha \varrho}}\,,\qquad \alpha = \sqrt{ \frac{ 2( \omega_0+2 )}{2\omega_0 +3} }\,,\qquad
a(t) = a_0\l(\frac{t}{t_0}\r)^{\frac{ 
2\omega_0(2-\gamma)+2}{3\omega_0\gamma(2-\gamma)+4}}\equiv 
a_{\ast}t^{\beta}\,,\nonumber
\ee
where $\gamma-1$ is the barotropic index of the equation of state for the cosmological 
perfect fluid. This metric is an explicit solution of Brans--Dicke theory, which is described by the action
\be
S_{BD}=\int d^4x \, \sqrt{-g} \left[ \phi R -\frac{\omega_0}{\phi} 
\, g^{\mu\nu} \nabla_{\mu}\phi \nabla_{\nu}\phi +\,{\cal 
L}^{(m)} \right] \,,\nonumber
\ee
where the effective gravitational coupling is proportional to the inverse of the scalar field $\phi(\varrho, t)$.  The constant $\omega_0$ 
is often called the Brans--Dicke parameter.

We would like to understand the dynamics of horizons (black hole and cosmological) in the Clifton--Mota--Barrow solution. 
Given that the solution is asymptotically  
Friedman--Lema\^itre--Robertson--Walker (FLRW) and 
dynamical, the most straightforward way  
to do so is to look for apparent horizons (note, however, potential caveats in the use for apparent horizon, as they depend on the foliation 
\cite{Wald:1991zz,Schnetter:2005ea}).  An apparent horizon (elsewhere dubbed ``trapping horizon''\cite{Hayward:1993wb})
is a  space- (or time-)like surface  defined as the closure 
of a surface foliated by marginally trapped
surfaces.

What follows is based on the analysis and results of Ref.~\refcite{Faraoni:2012sf}.

\section{Location of the apparent horizons}

The existence and location of the apparent horizons are  
determined by the  condition $g^{rr}=0$, where $r$ is the 
areal radius (not to be confused with the isotropic radius 
$\varrho$).  For the Clifton--Mota--Barrow solutions and 
for an expanding universe with $H \equiv \dot{a}/a>0$ this 
condition reads\cite{Faraoni:2012sf}
\be\label{main}
Hr^2-\frac{(\alpha -1)(\alpha +2)}{\alpha^2} \, m \, a(t) 
A(r)^{\frac{2(\alpha -1)( \alpha +1)}{\alpha}}-A(r)^{\alpha +1}r =0 \,,
\ee
where $H=\beta/t$ is the Hubble parameter corresponding to the scale factor $a(t)$.

The scalar curvature is 
singular in the limit $r\rightarrow0$, denoting the presence of a central singularity.
Once the variable $x\equiv \frac{m}{2\alpha \varrho}$ has been introduced, eq.~(\ref{main}) can be solved parametrically for the radius 
$r$ of the apparent horizon(s) and the time coordinate $t$, 
\bea
r (x) & = & a_{\ast} t^\beta \frac{m}{2\alpha} 
\, \frac{(1+x)^2}{x} \l(\frac{1-x}{1+x}\r)^{\frac{(\alpha -1)( \alpha 
+2)}{\alpha}}   \,,\nonumber\\
%&&\nonumber\\
t(x) &  =&\!\! \l\{ \!\frac{2\alpha}{m \, a_{\ast}\beta} \, 
\frac{x}{(1+x)^{\frac{2}{\alpha}(\alpha +1)}}\!\l[
(1-x)^{2/\alpha}  2x \, \frac{(\alpha -1)(\alpha +2)}{\alpha}  \, 
(1-x)^{-2(\alpha - 1)/\alpha }\r] \!\r\}^{\frac{1}{\beta-1}}\!. \nonumber
\eea

As a typical example, we plot the radius of the apparent 
horizon as a function of time (in adapted units) for the 
case $\omega_0=1$. The red (dashed) curve is 
for dust ($\gamma_D=1$), the green (solid) curve is
for both radiation and stiff matter ($\gamma_R=4/3$ and $\gamma_{SM}=2$), while the blue (dotted) one corresponds to a cosmological constant ($\gamma_\Lambda=0$). 
The initial behaviour (a unique, expanding apparent horizon) is common in all of these different configurations.
\begin{figure}[ht!]
\begin{center}
\parbox{2.4in}{\epsfig{figure=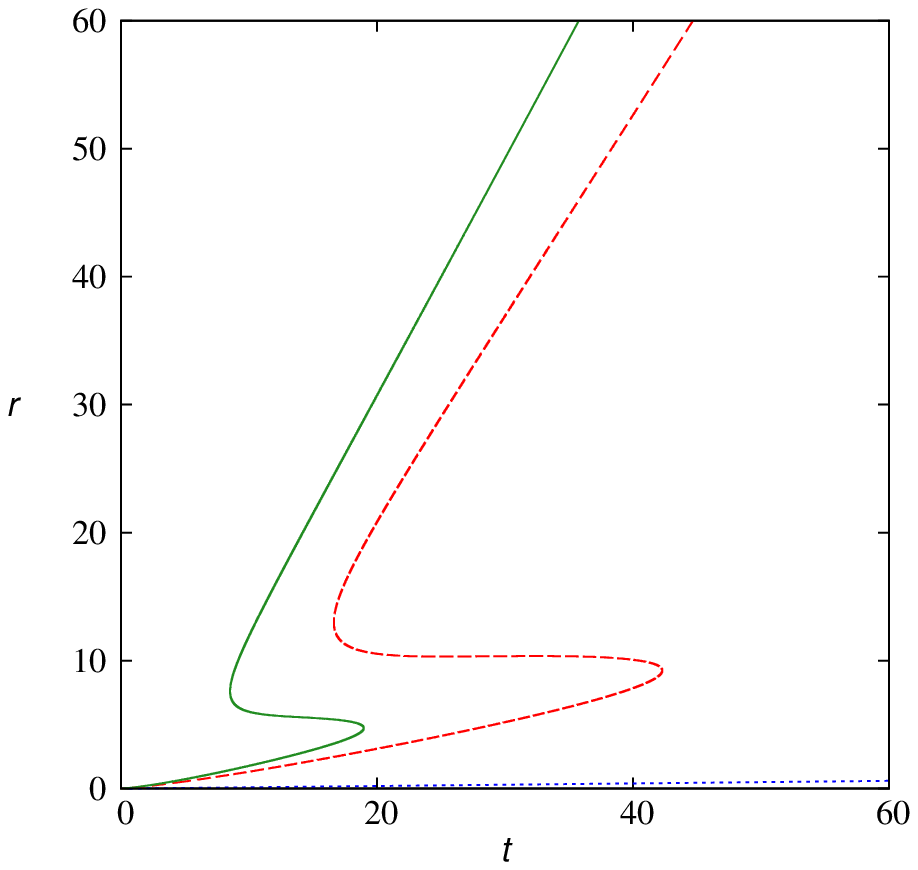,width=2.4in}}
\hspace*{4pt}
\parbox{2.4in}{\epsfig{figure=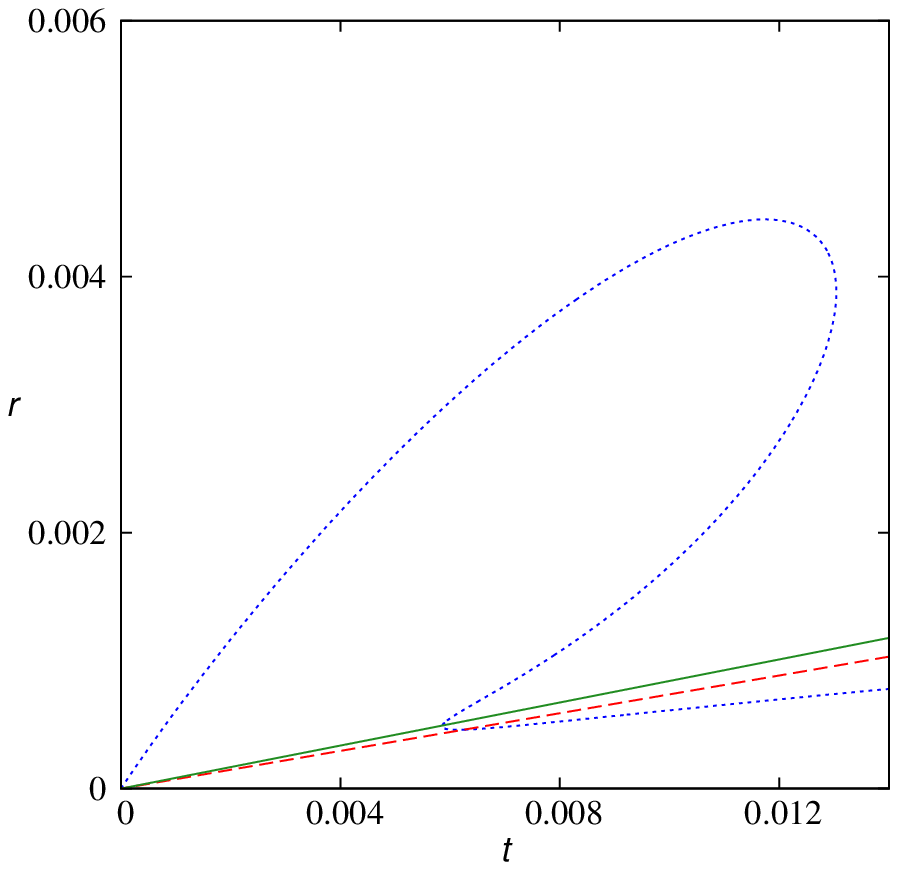,width=2.4in}}
\caption{Radius of the apparent horizon as function of time in adapted units, for $\omega_0=1$. The right panel focuses on the early time region of the plot.}
\label{fig1}
\end{center}
\end{figure}

For dust, radiation, and stiff matter, two 
further horizons appear at a certain time $t_*$, the outermost of which expands forever, while the 
other merges with the initial horizon, leaving behind a naked singularity covered by a cosmological horizon (cf. the contribution by V. Faraoni to these Proceedings, 
a similar solution by Husain, Martinez, and Nunez\cite{Husain:1994uj}, and its interpretation\cite{Faraoni:2012im}). The cosmological constant case is different in the fact 
that the two horizons appear inside the initial one, as shown in the right panel of Fig. \ref{fig1}.

The complete analysis of the Clifton-Mota-Barrow solution 
reveals a richer phenomenology\cite{Faraoni:2012sf} than 
just the previous example. In particular,  
the general relativity  limit $\omega_0\rightarrow\infty$  
turns out to correspond to a generalized McVittie 
metric\cite{Faraoni:2008tx}. 
In this case an initial naked singularity is then  covered by two appearing horizons. This is true for any equation of state of the cosmological fluid except
for the cosmological constant. For the latter  the 
reverse happens, that is, an expanding and a contracting horizon merge to leave behind a naked singularity.

We conclude with a warning: as already mentioned, apparent horizons depend on the spacetime slicings adopted, namely, specific slicings could imply the absence of 
apparent horizons even though the singularity is hidden by an event horizon. From this perspective, the issue of the presence of the naked spacetime singularity 
reported above becomes particularly delicate.
Nonetheless, looking for apparent horizons is probably the most straightforward probe for event horizon and it seems unlikely that slicing dependence seriously affects the results for a spherically symmetric 
foliation of a FLRW spacetime.

\section*{Acknowledgements}
VV is supported by FCT - Portugal through the grant SFRH/BPD/77678/2011.
VF would like to thank NSERC for 
financial support. TPS acknowledges financial
support provided under the Marie Curie Career Integration
Grant 	LIMITSOFGR-2011-TPS and the European Union's FP7 ERC Starting Grant "Challenging General Relativity" CGR2011TPS, grant agreement no. 306425.


\begin{thebibliography}{99}

  \bibitem{McVittie:1933zz} 
  G.~C.~McVittie,
  Mon.\ Not.\ Roy.\ Astron.\ Soc.\  {\bf 93}, 325 (1933).

\bibitem{Clifton:2004st} 
  T.~Clifton, D.~F.~Mota and J.~D.~Barrow,
  Mon.\ Not.\ Roy.\ Astron.\ Soc.\  {\bf 358}, 601 (2005).
  
\bibitem{Wald:1991zz} 
  R.~M.~Wald and V.~Iyer,
  Phys.\ Rev.\ D {\bf 44}, 3719 (1991).

\bibitem{Schnetter:2005ea} 
  E.~Schnetter and B.~Krishnan,
  Phys.\ Rev.\ D {\bf 73}, 021502 (2006).

\bibitem{Hayward:1993wb} 
  S.~A.~Hayward,
  Phys.\ Rev.\ D {\bf 49}, 6467 (1994).

\bibitem{Faraoni:2012sf} 
  V.~Faraoni, V.~Vitagliano, T.~P.~Sotiriou and S.~Liberati,
  Phys.\ Rev.\ D {\bf 86}, 064040 (2012).
  
\bibitem{Husain:1994uj} 
  V.~Husain, E.~A.~Martinez and D.~Nunez,
  Phys.\ Rev.\ D {\bf 50}, 3783 (1994).

\bibitem{Faraoni:2012im} 
  V.~Faraoni and A.~F.~Zambrano Moreno,
  Phys.\ Rev.\ D {\bf 86}, 084044 (2012).
  
\bibitem{Faraoni:2008tx} 
  V.~Faraoni, C.~Gao, X.~Chen and Y.~-G.~Shen,
  Phys.\ Lett.\ B {\bf 671}, 7 (2009).

\end{thebibliography}
\end{document}